\documentstyle[equations,12pt]{article}
\begin{document}
\begin{center}

{\Large \bf Enslaving random fluctuations in nonequlibrium systems} \\
\vskip 0.3 true in

{Mangal C. Mahato, T. P. Pareek and A. M. Jayannavar}\\

{Institute of Physics, Sachivalaya Marg, Bhubaneswar-751005, INDIA.}

\end{center}

\begin{abstract}
\par Several physical models have recently been proposed to
obtain unidirectional motion of an overdamped Brownian particle
in a periodic potential system. The asymmetric ratchetlike
form of the periodic potential and the presence of correlated
nonequilibrium fluctuating forces are considered essential 
to obtain such a macroscopic motion in homogeneous
systems. In the present work, instead, inhomogeneous systems are
considered, wherein the friction coefficient and/or temperature
could vary in space. We show that unidirectional motion can be
obtained even in a symmetric nonratchetlike periodic potential
system in the presence of white noise fluctuations.
We consider four different cases of system inhomogeneity
We argue that all these different models work under the same basic
principle of alteration of relative stability of otherwise
locally stable states in the presence of temperature
inhomogeneity. 

\end{abstract}


\eject

\newpage
\hspace {0.5in}  
{\bf I. Introduction} 
\par There has been much discussion recently[1-10] on how to model a
physical system that could extract work out of random
fluctuations without having to apply directly an obvious biased
force taking, in fact, cue from ---- or perhaps yearning to explain 
----- an experimentally observed phenomena[11] of predominantly
unidirectional motion of macromolecules (biological motors) 
along microtubules. We reason in the present work that system 
inhomogeneity may provide a clear and unifying framework to
approach the problem of macroscopic motion under discussion. 
Macroscopic unidirectional motion of a particle is not possible
thermodynamically in the prsence of equilibrium fluctuations.
However, such a motion can be obtained in an nonequilibrium
situation where the principle of detailed balance does not hold.
The existing popular models,[1-10] currently in the literature, mostly 
take the fluctuations to be nonequilibrium, that is, consider
nonwhite or at least nonguassian-white (colored) noise
together with a ratchetlike periodic system potential to aid
asymmetric motion of an overdamped Brownian particle.

\par The ratchetlike periodic system potentials, $V(q)$, 
obviously violate parity $V(q)\neq V(-q)$. 
For such a ratchetlike potential one can readily calculate
steady current flow $J(F)$ of a Brownian particle in the
presence of an external field $F$. It turns out that $J(F)$ is
not an odd function of $F$ and, in general, $J(F)\neq - J(-F)$. 
In other words, reversal of the external force may not lead to a reversed
current of the same magnitude in sharp contrast to the case of a
nonratchetlike (symmetric) periodic potential system where
$J(F)$= $-J(-F)$ follows. From this general observation, in a
ratchetlike potential, it can be easily concluded that on
application of a zero time averaged periodic field, say
$F$=$F_{0}sin\omega t$, one can obtain net unidirectional current.
Of course, the direction and
magnitude of the average velocity may depend, in a complicated way,
on the ratchetlike potential parameters, the thermal noise strength
as well as on the external field parameters, $F_{0}$ and $\omega$.
A careful tuning of
the relevant parameters may even result in the reversal of the
macroscopic current [2]. This is the basic physics behind 
some of the physical models used to obtain current rectification
in a periodic potential system.
There are models, however, that do not
use oscillating external fields. Instead, colored noise
of zero average strength----dichotomous,
Ornstein-Uhlenbeck, Kangaroo processes,..., [3-5]----- is used to drive the
Brownian particle to obtain macroscopic motion in a ratchetlike
potential system. There are further interesting models where the
potential barriers themselves are allowed to fluctuate, for
instance, with finite time correlations between two states under
the influence of a noise source. An example
being an overdamped Brownian particle subjected to a ratchetlike
periodic potential, where the ratchetlike saw-tooth potential is 
switched on to its full strength for time $\tau_{on}$ during which
the Brownian particle slides down the potential slope to the bottom of
the potential trench. At the end of $\tau_{on}$, the system is put in
the other ($off$) state during which the potential is set equal
to a constant (say = 0)
for an interval $\tau_{off}$ and the particle executes
force-free diffusive motion. At the end of $\tau_{off}$ the system is put back
in the $on$ state for interval $\tau_{on}$. This process of flipping of
states is repeated ad-infinitum. If $\tau_{off}$ is adjusted in such
a way that by the end of $\tau_{off}$ the diffusive motion just takes 
the particle out of the (now nonexistent) potential trench in the 
steeper slope direction (smaller distance) of the saw-tooth potential
but fails to do so in the gentler slope direction (larger distance),
the immediate next $on$ interval will take the particle to the adjacent 
trench minimum in the steeper slope side of the saw-tooth potential.
Repititon of such sequential flippings of states 
for a large number of times leads to a net unidirectional
macroscopic current of the Brownian particle. It should be noted that
a symmetrical nonratchetlike potential would, instead, have yielded symmetrical
excursions of the Brownian particle and, hence, no net unidirectional motion.
In this mechanism of obtaining net current of the 
Brownian particle the system is supplied
with the required energy externally to flip the system between the two states 
keeping the interval $\tau_{on}$ and $\tau_{off}$ fixed. There is a
lot of freedom to play around with the parameters $\tau_{on}$, $\tau_{off}$
, the parameters of the saw tooth potential, and the thermal (Brownian)
noise strength. And a judicious tuning of these parameters could 
even result in the reversal of the current. The flipping process could,
however, be effected also by a finite time-correlated 
fluctuating dichotomous force.
It should be noted that the former flipping 
proceses of definite $\tau_{on}$ and 
$\tau_{off}$ have been practically exploited in the particle separation
techniques, whereas the latter fluctuating flipping time process
has some appeal to natural processes.

The diversity of the models just
does not end here. There has been attempts too to obtain
macroscopic current with Gaussian white noise under nonratchetlike
symmetric periodic potential field as well but subjected to
temporally asymmetric periodic external fields [8-10]. We do not attempt
here, however, to give a review of the models considered. We
present, in the following, a framework to obtain macroscopic
motion in an inhomogeneous system with space dependent friction
coefficient and nonuniform temperature fields restricting
ourselves to periodic potential systems.

\par In all the models just mentioned [1-10] (The list is not
exhaustive.) the system was
taken to be homogeneous as far as the question of diffusivity
was concerned. However, in some of the works, earlier to the
ones eluded to so far, the nonuniformity of the diffusion constant
of the system was considered to yield macroscopic transport [12-13 ].
The diffusion coefficient could be space dependent or
state dependent and so the system may dissipate energy during
its time evolution differently at different space points.
Unlike homogeneous systems,
however, the physics of inhomogeneous systems has not been free
from controversies, such as, whether the equation
\begin{subequations}
\begin{eqalignno}
\frac{\partial P}{\partial t}=\frac{{{\partial}^2}}{\partial q^{2}}
{D(q)P},
\end{eqalignno}
or \\
\begin{equation}
\frac{\partial P}{\partial t}=\frac{\partial}{\partial q}{D(q)}
\frac{\partial P}{\partial q},
\end{equation}
\end{subequations}
\\
\noindent should be the correct form of diffusion equation. Nevertheless,
such controversies apart, B$\ddot u$ttiker [12] and also van
Kampen [13] have shown that one can expect macroscopic transport
of a Brownian particle in a periodic potential field when the
diffusion coefficient is also periodic with the same periodicity
but shifted by a phase difference other than 0 and $\pi$ with
respect to the periodic potential field. It should be noted that
the potential field is not required to be ratchetlike. The
system is rendered nonequilibrium by diffusion coefficient
inhomogeneity in the system and the "stationary state" of the
system is no longer goverened by the usual Boltzmann factor.

\par In reference [12], the source of inhomogeneity is not dwelt
into. However, there exist literature, that discuss at length the
thermodynamic origin of the possible inhomogeneity of the
diffusion coefficient [14]. The inhomogeneity could arise from
the nonuniformity of the friction coefficient and/or from the
nonuniformity of the temperature of the macroscopic sample. The
friction coefficient variation changes the rate of relaxation of
the system, whereas the temperature variation even alters the
relative stability of the otherwise locally stable states. The
idea of the change of relative stability of states due to temperature
inhomogeneity was advanced by Landauer [14]. This
idea has already been employed to show the possibility to
generate current in a closed ring without the application of any
external fields [13]. It is, however, important to understand the
contributions of the friction coefficient inhomogeneity and the
nonuniformity of temperature separately for, as stated earlier,
they influence the transport properties of the system in two
different ways. Moreover, to clarify the physical meaning of
various terms, in the theory, one needs to go beyond the
phenomenological description.

\par In reference [15], a microscopic treatment is given for the
derivation of the macroscopic equations of motion in an
inhomogeneous medium (space dependent friction coefficient and
spatially nonuniform temperature) starting from a microscopic
Hamiltonian of the system in contact with (phonon) heat bath(s).
Moreover, a proper overdamped limit of the Langevin equation of
motion in such an inhomogeneous medium is derived. 
A correct form of the corresponding Fokker-Planck equation is 
obtained and it is
explicitly shown that neither of the two forms of the diffusion
equation mentioned above [eq.(1)] is correct.
From this macroscopic
equation of motion one obtains an expression for the average
current which depends on the details of the potential field and
the inhomogeneities of the system. This microscopic treatment,
however, helps in understanding the functioning of the recently
proposed model of Maxwell's demon type information engine [16].

\par The Maxwell's demon type information engine in which the
particle is coupled to two
thermal baths at temperatures $T$ and $\overline T$ was investigated in
reference [16,17]. It has been shown
that the engine extracts work out of a nonequlibrium
environment ($T \neq \overline T$) by rectifying internal (white noise)
fluctuations and the Brownian particle, thus, acquires a nonzero
macroscopic velocity. A correct analytical expression for the
average macroscopic velocity of the Brownian particle is derived
and it is shown[17] that in an approximate range of physical
parameters the average velocity so acquired is similar to that
of a particle evolving in an inhomogeneous medium(see below).

\par As mentioned earlier, the nonuniformity of diffusion
coefficient can arise either because of the space dependence of
the friction coefficient $\eta(q)$ or that of the temperature
$T(q)$ or because of both. In physics such inhomogeneous systems
are not so uncommon. Examples of such systems include systems
with semiconductor junctions and a growing solid crystallite in the melt.
In the former system the inhomogeneity manifests itself
best when an electric current is flown through whereas in the
latter when the interface expands or contracts as a result of
heat current flow. We, however, discuss possibilities of
macroscopic current flow as a result of various kind of
inhomogeneities in a symmetric periodic potential
system. The first case we consider is when $\eta(q)$ and $T(q)$
are space dependent. 
By taking $\eta(q)$ and $T(q)$
periodic one gets a tilt in the potential field throughout the
sample as discussed phenomenologically in ref.[12],
resulting in a macroscopic current because of thermal
fluctuations. In this case the temperature inhomogeneity is
crucial and one can obtain current even when the friction
coefficient becomes uniform. Friction coefficient
inhomogeneity alone, however, does not generate macroscopic current. In
the second case, we consider a thermal particle in a system with
space dependent friction coefficient but subjected to 
external white noise fluctuations, and in the third 
case a thermal particle subjected to external space dependent
white noise is considered. Finally, we discuss the case of
a Brownian particle coupled to two thermal baths.
It should be noted, however, that in
all the cases that we are considering we do not require
the potential to be ratchetlike nor do we require the
fluctuation forces to be correlated in time to obtain
macroscopic current.

\par In section II we provide a derivation of the
macroscopic equation of motion in an inhomogeneous system [15] from a 
microscopic Hamiltonian of a Brownian particle interacting with a
(phonon ) heat bath. We, then, obtain proper Smoluchowski equation
from the derived Langevin equation of motion following the
prescription of Sancho et al [18]. We use, in Sec.III, this
overdamped equation of motion in an inhomogeneous system with space
dependent friction coefficient and nonuniform temperature field to 
obtain nonzero macroscopic current. In the same section we elaborate
three other possible cases of inhomogeneous systems where macroscopic
current could be possible. These are shown to be the special cases of a
general Maxwell's demon type information engine [16,17]. 
The section IV is devoted to conclusions.

II. {\bf Equation of motion in inhomogeneous systems}
\par We consider an inhomogeneous system where the inhomogeneity
could arise either because of the space dependence of friction
coefficient, or the nonuniformity of the temperature field or
because of the combined effect of both. The effect of the
nonuniformity of temperature or temperature gradient, however,
cannot be incorporated as a potential term in the Hamiltonian
formalism in sharp contrast to, for instance, the amenability of
incorporation of electric field gradient in the Hamiltonian of a
charged particle. We, therefore, incorporate the effect of
temperature inhomogeneity at the end directly into the equation 
of motion obtained
from the microscopic Hamiltonian suited to take care of the
space dependence of the friction coefficient.

{IIA. \bf  Equation of motion in a space dependent friction field.}
\par We consider a (subsystem) Brownian particle, of mass $M$,
described by a coordinate $Q$ and momentum $P$ moving in a potential
field $V(Q)$ of the system and being in contact with a
thermal(phonon) bath. The bath oscillators are described by
coordinates $q_{\alpha}$, momenta $p_{\alpha}$ and mass $m_{\alpha}$ 
with characteristic frequencies $\omega_{\alpha}$. We consider
the total Hamiltonian
\begin{equation}
H \: = \: \frac{P^2}{2M}+V(Q)+\sum_{\alpha}\left[\frac{p_{\alpha}^2}{2m_{\alpha}}
\: + \: \frac{m_{\alpha}\omega_{\alpha}^2}{2}
\left(q_{\alpha} \: - \: \lambda_{\alpha}\frac{A(Q)}
{m_{\alpha}\omega_{\alpha}^2}\right)^{2} \right],
\end{equation}
\noindent The interaction of the subsystem with the thermal bath
is through the linear (in $q$) coupling term
$\lambda_{\alpha}q_{\alpha}A(Q)$. From (2) one obtains the
following equations of motion.
\begin{subequations}
\begin{eqalignno}
{\dot Q} \: = \: \frac{P}{M} ,
\end{eqalignno}
\begin{eqalignno}
{\dot P} \: = \: - V^{\prime}(Q)+\sum_{\alpha}\lambda_{\alpha}
A^{\prime}(Q)\left[q_{\alpha} \: - \: \lambda_{\alpha}
\frac{A(Q)}{m_{\alpha}{\omega_{\alpha}^2}}\right],  
\end{eqalignno}
\begin{eqalignno}
{\dot q}=\frac{p_{\alpha}}{m_{\alpha}},
\end{eqalignno}
\begin{equation}
{\dot p_{\alpha}} \: = \: - m_{\alpha}{\omega_{\alpha}^2}q_{\alpha}
\: + \: \lambda_{\alpha}{A(Q)},
\end{equation}
\end{subequations}
\noindent where $A^{\prime}(Q)$ is the derivative of $A(Q)$ with respect
to $Q$. After solving (3c) and (3d) for $q_\alpha$ using the
method of Laplace transform and substituting its value in (3b),
we obtain
\begin{subequations}
\begin{eqalignno}
{\dot Q} \: = \: \frac{P}{M},
\end{eqalignno}
\begin{eqnarray}
{\dot P(t)} \: = \: - \: V^{\prime}(Q) \: - \: 
\sum_{\alpha}\frac{\lambda_{\alpha}^{2}
A^{\prime}(Q)}{m_{\alpha}\ {\omega_{\alpha}^2}}}
\int_{0}^{t}dx \: {\cos\omega_{\alpha}(t - t^{\prime}) A^{\prime}(Q)
{P(t^{\prime})\over M}  \nonumber\\
+A^{\prime}(Q) \sum_{\alpha} \lambda_{\alpha}
\left[ x_{\alpha}(0) 
\cos(\omega_{\alpha}t)
\: + \: {{\dot x}_{\alpha}(0) \over \omega_{\alpha}}
\sin(\omega_{\alpha}t) \right]  
+A^{\prime}(Q) \sum_{\alpha} \frac{A(Q_0) \lambda^{2}_{\alpha}}
{m_\alpha \omega^{2}_\alpha} \cos(\omega_\alpha t). \yesnumber
\end{eqnarray}
\end{subequations}
\noindent Here $Q_0$ is the initial value of the particle
co-ordinate $Q$ and $x_\alpha (0)$ and $\dot x_\alpha (0)$ are
the initial co-ordinates and velocities, respectively , of the
bath variables.
The second term in the right hand side of equation (4b)
depends on the momenta at all times
previous to $t$. At this stage Markovian limit is imposed so that
\begin{equation}
g(t - t^{\prime}) \: = \: \sum_{\alpha}\frac{{\lambda_{\alpha}^{2}}} 
{m_{\alpha}\omega_{\alpha}^2}\;
\cos{\omega_{\alpha}(t - t^{\prime})} = 2\eta \delta(t-t^{\prime}).
\end{equation}
\noindent The equation (5) follows readily from the well known
Ohmic spectral density distribution for the bath oscillators, i.e., 
\begin{equation}
\rho(\omega) \: = \: \frac{\pi}{2} 
\sum_{\alpha} \frac{\lambda^{2}_{\alpha}}{m\omega_{\alpha}}
{\delta(\omega - \omega_{\alpha})}= 
\eta\omega{e^{-{\frac{\omega}{\omega_{c}}}}}. 
\end{equation}
\noindent where $\omega_c$ is an upper cutoff frequency set by
the oscillator spectrum of the thermal bath.
The Markovian approximation (5) has the effect of 
neglecting the transient terms involving the initial coordinate $Q_0$, in
the equation of motion, for long time behaviour [19]. 
In other words, the equation should well describe
the motion of the Brownian particle in time scales $t >$
$\omega_{c}^{-1}$.
The equation of motion thus assumes the form
\begin{subequations}
\begin{eqalignno}
{\dot Q} \: = \: \frac{P}{M},
\end{eqalignno}
\begin{equation}
{\dot P} \: = \: - V^{\prime}(Q)  -  \frac{\eta}{M}
[A^{\prime}(Q)]^2{P} + A^{\prime}(Q)\ f(t), 
\end{equation}
\end{subequations}
\noindent where 
\begin{equation}
f(t) \: = \: \sum_{\alpha}\lambda_{\alpha}\left[q_{\alpha}(0)
\cos(\omega_{\alpha}t)
\: + \: \frac{\dot q_{\alpha}(0)}{\omega_{\alpha}} 
\sin(\omega_{\alpha}t) \right].
\end{equation}

\noindent The force $f(t)$ is fluctuating in character because of
the associated uncertainties in the initial conditions
$q_{\alpha}(0)$ and ${\dot q_{\alpha}(0)}$ of the bath variables.
However, as the thermal bath is characterized by its temperature
$T$, the equilibrium distribution $P_{eq}(q_{\alpha}(0),\dot q_{\alpha}(0))$
of bath variables is given by
the Boltzmannian form 

$$ \hskip 1.4 in P(q_{\alpha}(0),\dot q_{\alpha}(0))= \frac{1}{Z}
\prod_{\alpha} e^{- \frac{1}{2k_{B}T}
\left( m_{\alpha} \dot q_{\alpha}^{2}(0)
+m_{\alpha} \omega_{\alpha}{^2}  q_{\alpha}{^2}(0) \right)},
\hskip 1.38 in (8a) $$

\noindent where $Z$ is the partition function. Using equation (8a)
and (6) one can easily compute the statistical properties of the
fluctuating force $f(t)$. It is Gaussian with
\begin{subequations}
\begin{eqalignno}
\langle f(t) \rangle = 0,
\end{eqalignno}
\noindent and \\
\begin{eqalignno}
\langle f(t)f(t^{\prime}) \rangle = k_{B}T\ g(t - t^{\prime})
 = 2k_{B}T\eta \delta(t - t^{\prime}).
\end{eqalignno}
\end{subequations}
\noindent It should be noted that the effect of the interaction
term $\lambda_{\alpha}q_{\alpha}A(Q)$ in the Hamiltonian (2) is
to introduce a friction term and a fluctuating term $f(t)$ in the
equation of motion (7b). Moreover, $A^{\prime}(Q)$=constant
corresponds to a uniform friction coefficient. We redefine,
${[A^{\prime}(Q)]^2}\eta = \eta(Q)$ and $\frac{f(t)}
{\sqrt T} \rightarrow f(t)$ , and put $M=1$, in (7) to obtain,
\begin{subequations}
\begin{eqalignno}
{\dot Q} = P,
\end{eqalignno}
\begin{eqalignno}
{\dot P} = - V^{\prime}(Q) - \eta(Q)P + \sqrt{k_{B}{T}{\eta(Q)}} f(t),
\end{eqalignno}
\end{subequations}
\noindent with
\begin{subequations}
\begin{eqalignno}
\langle f(t) \rangle = 0,
\end{eqalignno}
\noindent and \\
\begin{eqalignno}
\langle {f(t)f(t^{\prime})} \rangle = 2 \delta(t - t^{\prime}).
\end{eqalignno}
\end{subequations}
\noindent From eqs.(9) it follows that the derived Langevin
equation of motion (10b) of a Brownian particle, in a system with
space dependent friction $\eta(Q)$ but at constant uniform
temperature $T$, is internally consistent and obeys fluctuation-dissipation
theorem. We now proceed to incorporate the effect of
space dependence of temperature, in a thermally nonuniform
system, into the Langevin equation of motion by assuming that
the Brownian particle comes in contact with a continuous
sequence of independent temperature baths as its coordinate $q$
changes in time. (For notational simplicity, we replace the
coordinate $Q$ and momenta $P$ by the corresponding lower case
letters $q$ and $p$, respectively, reserving $P$ for probability
distribution.) 

IIB. {\bf  Equation of motion in a space dependent friction and
temperature field}
\par We consider each space point $q$ of the system to be in
equilibrium with a thermal bath characterised by temperature
$T(q)$. Also, it should be noted that one could take $\eta(q)$ to
be constant piecewise along $q$ , and in each piece of these $q$
segments eq.(10b) would correspond to an equation of motion with
the constant friction
coefficient but with the same statistical character 
of $f(t)$ (11a-11b) in all $q$ intervals. Let us discretize
the system, for the sake of argument, into segments $\Delta{q}$
around $q$ and represent them by indices $i$. Let us further assume
that each segment is connected to an independent thermal bath at
temperature $T_i$ with corresponding random forces $f_{i}(t)$ so
that the equation of motion (10b), in the segment $i$, will have
the last term $\sqrt{k_{B}T_{i}{\eta(Q)}}f_{i}{(t)}$. As the two
different segments $i$ and $j$ are each coupled to an independent
tempearture bath we have $\langle
f_{i}(t)f_{j}(t^{\prime})\rangle = 2 \delta_{ij}\delta(t - t^{\prime})$.
Because $f(t)$ is $\delta$ correlated in time, as the particle
evolves dynamically the fluctuation force $f_{i}(t)$ experienced
by the Brownian particle while in the space segment $i$ at time $t$
will have no memory about the fluctuating force experienced by
it at some previous time $t^{\prime}$ while in the space segment
$j \neq i$. The space-dependent index $i$ in $f_{i}(t)$,
therefore, can be ignored and the equation of motion becomes
local in time as well as in space. Therefore, in the continuum
limit, the stochastic equations of motion of the Brownian
particle, in an inhomogeneous medium with space dependent
friction and nonuniform temperature, acquire the simple forms
\begin{subequations}
\begin{eqalignno}
{\dot q} = p,
\end{eqalignno}
\begin{eqalignno}
{\dot p} = - V^{\prime}(q) - \eta(q)p + \sqrt{k_{B}T(q)\eta(q)} f(t),
\end{eqalignno}
\noindent with
\begin{eqalignno}
\langle f(t)f(t^{\prime}) \rangle = 2 \delta(t - t^{\prime}).
\end{eqalignno}
\end{subequations}

IIC. {\bf The Smoluchowski Equation}
\par From eq.(12b) one can readily write down the Fokker-Planck
equation or the Kramers equation for the full probability
distribution $P(q,v,t)$. However, in most of the practical
situations the marginal probability distribution $P(q,t)$ for the
variable $q$ alone suffices to describe the motion of the
Brownian particle. This probability distribution $P(q,t)$ can be
obtained in the overdamped limit of the Langevin equation (12b)
which is valid on time scales larger than the inverse friction $\eta^{-1}$.
In other words in the overdamped case the fast
variable, velocity $v$, is eliminated from the equation of motion. 
In the case of
homogeneous systems one simply puts ${\dot p} = 0$ in eq.(12b) to obtain the 
overdamped Langevin equation. However, in case of inhomogeneous systems, the
above method of adiabatic elimination of fast variables 
does not work, and leads to unphysical equilibrium distribution. 
The proper prescription for the elimination of
fast variables has been given in Ref.[18] 
for systems with space dependent friction. 
The method retains all terms upto order $\eta^{-1}$ 
and the resulting overdamped Langevin equation yields 
physically valid equilibrium distribution. We,
therefore, apply the same prescription to obtain the overdamped 
Langevin equation
of motion in an inhomogeneous system with space dependent friction $\eta(q)$ 
 and nonuniform temperature field $T(q)$. We obtain,
\begin{equation}
{\dot q} = - \frac{V^{\prime}(q)}{ \eta(q)} -
\frac{k_{B}}{2 {[\eta(q)]}^2}
\left[T(q) \eta^{\prime}(q) + \eta(q)T^{\prime}(q) \right] + 
\sqrt{ \frac{k_{B}T(q)}{\eta(q)}}f(t),
\end{equation}
\noindent with \\
\begin{equation}
\langle f(t)f(t^{\prime}) \rangle \: = \: 2 \delta(t - t^{\prime}).
\end{equation}

\noindent Using van Kampen Lemma[20] and the Novikov's
theorem[21] we obtain the
corresponding Fokker-Planck equation as

\begin{equation}
{\partial P(q,t) \over \partial t} \: = \: {\partial \over \partial q}
{1 \over \eta(q)}\left[{\partial \over \partial q}k_{B}T(q)P(q,t) \: + \:
V^{\prime}(q) P(q,t)\right].
\end{equation}

\noindent Eq.(15) is the Smoluchowski equation for an 
overdamped Brownian particle
moving in an inhomogeneous system with space dependent friction and 
nonuniform temperature. It should be noted that eq.(15) 
gives the correct form of
diffusion equation instead of either of the two forms mentioned in eqs.(1).
It is clear that the temperature and the friction coefficients influence the 
particle motion in a qualitatively different fashion and they cannot be 
plugged
together to get effective diffusion coefficient to satisfy either of the forms
of eq.(1). In the next section we discuss how the
system inhomogeneity can 
help maintain a macroscopic unidirectional current.

III. {\bf Macroscopic motion obtained from inhomogeneous systems}
\par We consider inhomogeneous systems where the inhomogeneity could be an
internal property of the system or it could be imposed externally. As mentioned 
earlier we consider four cases where macroscopic motion can be obtained.

IIIA. {\bf Macroscopic motion in an inhomogeneous 
system with space dependent 
friction and nonuniform temperature}
\par When the system is bounded at $q \rightarrow \pm \infty$, i.e.,
$V \rightarrow \infty $ as $q \rightarrow \pm \infty$, the system
attains steady (stationary) state with zero probability current.
In such a situation, we can calculate the steady state
probability distribution $P_{s}(q)$, from the Smoluchowski
equation (15), by setting the probability current
\begin{equation}
{1 \over \eta(q)}\left[V^{\prime}(q)P(q,t) + {\partial \over \partial q} 
k_{B}T(q)P(q,t)\right]
\end{equation}

\noindent equal to zero, as 
\begin{equation}
P_{s}(q) = {\it N} e^{-\psi(q)},
\end{equation}
\noindent where\\
\begin{equation}
\psi(q) \: = \: \int^{q} \left(V^{\prime}(x) + 
kT^{\prime}(x) \over k_{B}T(x)\right) dx,
\end{equation}\\
\noindent and {\it N} is a normalization constant. 
It is very clear from the expression, eq.(18), for $\psi (q)$
that the peaks of $P_{s}(q)$ are determined not by the minima
of $V(x)$ alone but are determined as a combined effect with
$T(x)$. $P_{s}(q)$ may even peak at positions which would be
quite less likely to be populated in the stationary situations
for uniform temperature, $T(x)$ = $T$ condition.
In this respect, nonequlibrium situations appear strange but are
quite common in biological systems where, for example, otherwise
less likely ion channels are, in some situations, found to be
more active for ionic transport. Recently such nonequlibrium
behaviour in biological systems have been theoretically
attributed to the effect of nonequlibrium fluctuations and the
process has been termed as kinetic focussing[22]. Moreover, it
should be noted that the relative stability of two states of a
system with nonuniform temperature field is not determined by
the local function $V(q)$ but by the entire pathway through a
continuous sequence of intervening states between the two states
under comparison. The
temperature variation may modify the kinetics of these
intervening states drastically and hence their contribution
towards the relative stability will be substantial even when they are
sparsely populated.
It should further be noted 
that $\psi(q)$ is not determined by $\eta(q)$ as it should be.
Moreover, the functional form of $\psi(q)$ is similar to
$\int^{q}{v(x) \over D(x)}dx$, of course, in this case $V^{\prime}(q)$ 
has been augmented by a compensating force k$T^{\prime}(q)$.
$v(q) = \eta^{-1}(q)[V^{\prime}(q)+kT^{\prime}(q)]$ is the drift 
velocity and D(q) = $\eta^{-1}(q)k_{B}T(q)$ is the effective diffusion
coefficient.

\par So far we have not assigned any functional form to $V(q), T(q)$
and $\eta(q)$.
In ref.[12] it is shown that at least in one case the system can
generate nonzero probability current, namely, when both $V(q)$
and $D(q)$ are periodic with same periodicity but having a 
phase difference other than 0 and $\pi$. 
In our present problem if we assume $V(q)$, $T(q)$ and $\eta(q)$
to be
periodic functions with periodicity, say, $2\pi$ then the probability current is
given by
\begin{equation}
J \: = \:  {{1 - e^{-\delta}} \over {\int_{0}^{2\pi}dy e^{-\psi(y)}\:
\int_{y}^{y+2\pi}dx {e^{\psi(x)} \over D(x)}}} ,
\end{equation}
\noindent where $\delta = \psi(q) - \psi(q+2\pi)$ determines the
effective slope of a generalized potential $\psi(q)$ and 
hence $\delta$ being $+$ or $-$ve  determines
the direction of current. It is obvious from the expression for  
$\delta$ that the phase difference $\phi$, between $V(q)$ and $T(q)$, alone
determines the direction of current. The net unidirectional current remains
nonzero (finite) except when $\phi$ is an integral mutiple of $\pi$
($\phi = n \pi$ corresponds to zero effective slope of the generalized 
potential). It should further be noted that the variation of $\eta(q)$ does not
determine the direction of current but does affect the magnitude of
current. A periodic variation of $\eta(q)$ and $V(q)$ but uniform
$T(q)$ will yield no unidirectional current.
For $V(q) = V_{0} (1 - cos(q))$ and $T(q) = 
T_{0}(1 - {\alpha \cos(q-\phi)})$, with $0 < \alpha < 1 $ 
 (for positive temperature) $\delta$ turns out to be
$\frac{2\pi V_{0} \sin\phi}{k_{B}T} 
\left[ \frac{1}{\sqrt{1- \alpha^2}}-1 \right]$ which is definitely
nonzero for $\phi \neq n\pi$, $n = 0,\pm 1 , \pm2$,.......
Thus $\phi$ alone determines the direction of 
nonzero current $J$. In this case the periodic 
variation of temperature plays the crucial
role and may yield current even when $\eta(q) = \eta_{0}$ = constant.
We now consider cases, where $\eta(q)$ plays a decisive role.

IIIB. {\bf Macroscopic motion in an inhomogeneous system with space dependent
friction in the presence of an external parametric white noise}
\par Unlike the case considered in subsection IIIA, where the
overdamped Brownian particle experiences a fixed (in time) local
(nonuniform) temperature profile $T(q)$ during its sojourn
$q(t)$ for all $t$,
we consider, in this subsection, a system with uniform temperature
$T(q)=T$ but a spatially varying $\eta(q)$.
The Langevin equation of motion is given by
\begin{equation}
{\dot p} = -V^{\prime}(q) - \eta(q) p + \sqrt{k_{B}T \eta(q)} f(t)
\end{equation}
\noindent and the corresponding overdamped equation is
\begin{equation}
{\dot q} = - \frac{V^{\prime}}{\eta(q)} - \frac{k_{B}T\eta^{\prime}(q)}
{2[\eta(q)]^2} + \sqrt{\frac{k_{B}T}{\eta(q)}} f(t),
\end{equation}
\noindent with
$$\langle f(t) \rangle = 0,$$ and
$$\langle f(t)f(t^{\prime}) \rangle = 2 \delta(t - t^{\prime}).$$
We, now, subject the system to an external parametric additive
white noise fluctuating force $\xi(t)$, so that the equation of
motion becomes
\begin{equation}
{\dot q} = - \frac{V^{\prime}}{\eta(q)} - \frac{k_{B}T\eta^{\prime}(q)}
{2[\eta(q)]^2} + \sqrt{\frac{k_{B}T}{\eta(q)}} f(t) + \xi(t),
\end{equation}
\noindent with
$$\langle \xi(t) \rangle = 0,$$
$$\langle \xi(t) \xi(t^{\prime}) \rangle = 2 \Gamma 
\delta(t - t^{\prime}), \eqno(22a)$$
\noindent where $\Gamma$ is the strength of the external white noise $\xi(t)$.
We can immediately write down the corresponding 
Fokker-Planck (Smoluchowski) equation
\begin{equation}
\frac{\partial P}{\partial t} \: = \:  \frac{\partial}{\partial q}
\left[\   \left\{\frac{V^{\prime}(q)}{\eta(q)} \right\} P 
 + \left\{\frac{k_{B}T}{\eta(q)} + \Gamma \right\}
\frac{\partial P}{\partial q} \right].
\end{equation}
\noindent For periodic functions $V(q)$ and $\eta(q)$, with
periodicity $2\pi$, one obtains unidirectional current following
earlier procedure using equation (23). The resulting expression
for current $J$ takes the same functional form as given in
eq.(19) where $\psi(q)$ is now given by 
\begin{equation}
\psi(q) = \int^{q} dx \frac{ V^{\prime}(x)}{k_{B}T+\Gamma \eta(x)}
\end{equation}
\noindent and the effective diffusion coefficient \\
\begin{equation}
D(q) = (k_{B}T + \Gamma \eta(q))/\eta(q),
\end{equation} 
\noindent with 
$$\delta = \psi(q) - \psi(q+2\pi). $$\\
For V(q) = $V_{0}(1-\cos(q))$ and $\eta(q) = \eta_{0} (1 -
\alpha \cos(q - \phi))$
$\delta$ turns out to be equal to
$ \frac{2\pi V_{0} \sin\phi}{ \alpha \eta_{0}}
\left[ \frac{k_{B}T+\eta_{0}}{\sqrt{(k_{B}T+\eta_0)^{2}-
(\eta_{0}\alpha)^2}}-1 \right].$
As earlier the direction of current is determined by the phase
difference $\phi$ between the periodic functions $V(q)$ and $\eta(q)$.

\par It is important to notice that there is no way 
one could obtain macroscopic
current in the absence of the external white noise $\xi(t)$. 
This case , however, is similiar in
essence to the previous case of nonuniform temperature. 
In the present situation the overdamped Brownian particle is
subjected to an external parametric random noise. The noise
being externally imposed, the system always absorbs energy
(without the presence of corresponding loss factor) [23].
Also, the overdamped
particle moves slowly wherever the friction coefficient $\eta(q)$ is 
large and the possibility of absorption of energy from the external white
noise at those elements $q$ of the system, therefore, is
correspondingly large. Thus, the effective temperature $T(q)$ of
the system is given by $k_{B}T+\Gamma \eta(q)$,
which modulates as $\eta(q)$ varies and
hence the macroscopic current results as in the case IIIA.

IIIC. {\bf Macroscopic motion in a 
homogeneous system but subjected to an external
parametric space dependent white noise}

\par The overdamped Langevin equation is ,
\begin{equation}
{\dot q} = -\frac{V^{\prime}(q)}{\eta} + \sqrt{\frac{k_{B}T}{\eta}} f(t),
\end{equation}
\noindent with $\langle f(t) \rangle$ = 0 and 
$\langle f(t)f(t^{\prime}) \rangle$
 = 2  $\delta(t - t^{\prime})$.
\noindent Eq.(26) obeys fluctuation-dissipation theorem and
hence in the absence of any external bias potential there can be
no net current irrespective of the form of the periodic
potential $V(q)$. We now subject the system to an external
multiplicative Gaussian white noise fluctuation. The
corresponding overdamped Langevin equation is given by
\begin{subequations}
\begin{eqalignno}
{\dot q} = -\frac{V^{\prime}(q)}{\eta} + \sqrt{\frac{k_{B}T}{\eta}} f(t)
+g(q) \xi(t),
\end{eqalignno}
\end{subequations}
\noindent where $g(q)$ is an arbitrary function of $q$ and
$\xi(t)$ is a white noise with 
$$\langle \xi(t) \rangle = 0,$$ and
$$\hskip 1.9 in \langle \xi(t) \xi(t^{\prime}) \rangle = 2\Gamma
\delta(t-t^{\prime}). \hskip 2.5 in (27b) $$ \\
The associated Fokker-Planck equation can
be immediately written down as
\begin{equation}
\frac{\partial P}{\partial t}  =  \frac{\partial}{\partial q}
\frac{V^{\prime}(q)}
{\eta}P+ \frac{k_{B}T}{\eta} \frac{\partial}{\partial q^2}P+
\Gamma \frac{\partial}{\partial q}g(q) \frac{\partial}{\partial q}g(q)P 
\end{equation}
\noindent Now, if we assume $V(q)$ and $g(q)$ to be periodic functions
with periodicity $2\pi$, the net unidirectional current can be obtained
and is given by eq.(19), with 
\begin{equation}
\psi(q)  =  \int^{q} dx \frac{V^{\prime}(x)  +  
\eta \Gamma g(x)g^{\prime}(x)}
{k_{B}T  +  \eta \Gamma g^{2}(x)},
\end{equation}
\noindent and the effective difussion coefficient,
$$ D(q)=\frac{k_{B}T+\eta \Gamma g^{2}(q)}{\eta}.$$
For the specific form of the
periodic functions $$V(q)=V_{0}(1-cosq),$$ and \\
$$g(q)= \sqrt{g_{0}(1- \alpha cos(q- \phi))},$$ we obtain, 
$$\delta= \frac{2\pi V_{0} \sin\phi}{\eta \Gamma g_{0} \alpha}
\left[ \frac{k_{B}T+\eta \Gamma g_{0}}{\sqrt{(k_{B}T+\eta \Gamma
g_{0})^{2} - (\eta \Gamma g_{0})^{2}}} -1 \right] .$$ 
The phase $\phi$ being +ve or -ve determines the
sign of $\delta$ and consequently direction of the current $J$ (eq.(19)).

\par It should be noted that, as in case IIIB, the overdamped Brownian
particle experiences an effective space dependent temperature 
$T(q)= k_{B}T+\eta \Gamma  [g(q)]^{2}$. 
The first case (IIIA) corresponds to a system which is intrinsically 
nonequilibrium and requires an internal mechanism 
such as the generation of latent heat
at the interface in first order transitions to maintain the temperature
profile $T(q)$. The other two cases (IIIB and IIIC) are, 
however, supplied with energy
externally via the externally applied white noise. And finally, we
consider a case where the Brownian particle is subjected to two 
thermal baths. 

IIID. {\bf Macroscopic motion in an inhomogeneous system under the
action of two thermal (noise) baths}
\par We now consider the situation in which the system is in contact
with an additive thermal noise bath at temperature $T$ and a multiplicative
thermal noise bath at temperature $\overline T$. The corresponding equation of
motion of the Brownian particle can be derived from a microscopic Hamiltonian
and is given by [16]
\begin{equation}
M{\ddot q}=-V^{\prime}(q)- \Gamma(q) \dot q + \xi_{A}(t)
+\sqrt{f(q)}\xi_{B}(t),
\end{equation}
\noindent $\xi_{A}(t)$ and $\xi_{B}(t)$ are two independent Gaussian white
noise fluctuating forces with statistics,
\begin{subequations}
\begin{eqalignno}
\langle \xi_{A}(t) \rangle\;=\;0,
\end{eqalignno}
\begin{eqalignno}
\langle \xi_{A}(t) \xi_{A}(t^{\prime})
\rangle\;=\;2\Gamma_{A}k_{B}T\delta(t-t^{\prime}),
\end{eqalignno}
\end{subequations}
\noindent and \\
\begin{subequations}
\begin{eqalignno}
\langle \xi_{B}(t) \rangle\;=\;0,
\end{eqalignno}
\begin{equation}
\langle \xi_{B}(t) \xi_{B}(t^{\prime})
\rangle\;=\;2\Gamma_{B}k_{B}{\overline T}\delta(t-t^{\prime}),
\end{equation}
\end{subequations}
\noindent where, $T$ and ${\overline T}$ are temperatures of the two baths A
and B, respectively. It should be noted that, $\xi_{A}(t)$ and
$\xi_{B}(t)$ represent internal fluctuations and together satisfy the
fluctuation-dissipation theorem $\Gamma(q)=\Gamma_{A}+\Gamma_{B} f(q)$. 
The bath B is associated with a space dependent friction coefficient
$f(q)$. When the two temperatures $T$ and $\overline T$ become
equal the system will be in equilibrium and no net current can flow.
By making $T$ and $\overline T$ different the system is rendered
nonequilibrium and one can extract energy at the expense of increased
entropy. The system, thus, acts as a Maxwell's-demon type information
engine which extracts work by rectifying internal fluctuations.
In ref. [16] an expression for current is obtained in the overdamped limit. 
The overdamped limit of the Langevin equation is taken by setting the 
left hand side of eq.(30) equal to zero. This procedure of obtaining 
overdamped limit is not correct as explained in section I. Following
the procedure of ref.[18] the correct Fokker-Planck equation in the  
overdamped limit is given by [17]
\begin{eqnarray}
\frac{\partial P}{\partial t}=
 \frac{\partial}{\partial q}
\left\{ \frac{V^{\prime}(q)}{\Gamma(q)} {P}+\frac{T\Gamma_{A}}{\Gamma(q)}
\frac{\partial}{\partial q} \frac{P}{\Gamma(q)} 
+{\overline T}{\Gamma_{B}} \frac{\sqrt{f(q)}}{\Gamma(q)}
\frac{\partial}{\partial q} \frac{\sqrt{f(q)}}{\Gamma(q)} {P} \right .\nonumber\\ 
\left .+{\overline T}{\Gamma_{B}} \frac{(\sqrt{f(q)})^{\prime} \sqrt{f(q)}}
{[\Gamma(q)]^{2}} {P} \right\}.
\end{eqnarray}
\noindent For periodic functions $V(q)$ and $f(q)$ with periodicity
$2\pi$ the noise induced transport current $J$ is given by eq.(19), 
where, now
\begin{equation}
\psi(q)=\int^{q}\left\{\frac{V^{\prime}(x)\Gamma(x)}{T\Gamma_{A}+
{\overline T}{\Gamma_{B}}f(x)}+\frac{(\overline T - T)}{\Gamma(x)}
\frac{\Gamma_{A}\Gamma_{B}f^{\prime}(x)}{(T\Gamma_{A}
+{\overline T}\Gamma_{B}f(x))}\right\}dx,
\end{equation} 
\noindent and  $D(q)=\frac{T\Gamma_{A}+{\overline T}\Gamma_{B}f(q)}
{(\Gamma(q))^2}$ and
$\delta=\psi(q)-\psi(q+2\pi).$
\par As in earlier cases, taking specific periodic forms
of $V(q)=V_{0}(1- \cos q)$ and $f(q)=f_{0}(1 -\alpha \cos(q -
\phi))$, the exponent $\delta$ in eq.(19) for current is
obtained as
\begin{equation}
\delta = \left( 1 - \frac{T}{\overline T} \right)
\frac{2\pi V_{0} \sin\phi}{{\overline T} \Gamma_{B} f_{0}
\alpha} \left[ \frac{T\Gamma_{A}+{\overline T} \Gamma_{B} f_{0}}
{\sqrt{(T\Gamma_{A}+{\overline T}\Gamma_{B} f_{0})^2 -
({\overline T}\Gamma_{B} f_{0} \alpha)^2}} -1 \right]
\end{equation}  
It is clear from the expression for $\delta$ that, again as in
earlier cases (IIIA-C), the phase difference $\phi$ between
$V(q)$ and $f(q)$ determines the direction of current $J$. It is
to be noted that, the current will flow in one direction if 
$ T >$ $\overline T$ and will flow in the opposite if $T < \overline
T$ for given $\phi$. Thus, the system acts like a Carnot engine
which extracts work by making use of two thermal baths at
different temperatures ($T \neq \overline T$). Moreover $\delta$
vanishes when $f(q)$ becomes space independent constant $f_0$,
i.e., when $\alpha = 0$, and the current $J$ becomes zero.
It should be noted further 
that when the amplitudes of $f(q)$ and $f^{\prime}(q)$ are
small compared to the amplitude of $V(q)$, the problem
turns out to be equivalent to a particle moving in a spatially varying
temperature field, $T(q)$=$(T\Gamma_{A}+{\overline
T}\Gamma_{B}f(q))/\Gamma(q)$ and, as discussed in section IIIA, such
a nonuniform temperature field yields net unidirectional current.

IV. {\bf Summary and Discusssion} 
\par Transport in a nonequlibrium periodic system has become, in
recent times, a field of very active research. We have just
tried, in the beginning of this work, only to enumerate various
working ideas to build a plausible model of thermal
ratchet. The brief enumeration is, of course, not complete. The
models are being gradually refined and simplified to be close
either to the experimental reality or to invent techniques
to be useful in practice. For example, there are attempts to
show that one can obtain macroscopic current in a symmetric
periodic potential system with the application of (1) zero averaged
external white shot noise with Poissonian waiting time distribution [24]
, (2) zero averaged but otherwise temporally asymmetric fluctuations [10],
or, (3) zero averaged asymmetric noise form with one large kick in one
direction and two smaller kicks of half the strength of the former
in the opposite direction [25]. Moreover, it has further been 
reported to obtain macroscopic motion with the use of two baths 
one thermal other athermal (time correlated) in a ratchetlike 
potential system [26]. In these situations even current reversal 
is possible. It has further been reported that with the application
of suitable voltage fluctuations a voltage-sensitive macromolecule
can be put into a desired (but otherwise energetically
unfavourable)  kinetic substate [22]. These vast new 
developements in the study of transport phenomena could be 
but interesting variants, in principle, of a simple unified
theoretical framework of inhomogeneous nonequlibrium systems.
Transport in inhomogeneous nonequilibrium systems has attracted
attention since long. We have presented in Sec.IIIA a microscopic approach
to obtain macroscopic equation of motion in such systems. 

\par In the situations considered, in sections IIIB and IIIC, 
the system is subjected to external fluctuations violating the fluctuation-
dissipation theorem.
Also, these two cases, in a sense, are
physically equivalent to having a spatially varying temperature 
field as considered in section IIIA. However, in the
situation IIID both the temperature baths 
were internal parts of the system and
the system is subjected to a space dependent friction at
temperature $\overline T$. 
Also in the limit
of small friction field modulation amplitude the situation
corresponds to a spatially varying temperature field. These
observations seem to suggest that the case of inhomogeneous
systems with spatially varying temperature field provides a
general paradigm to obtain macroscopic current and several
variants considered to model fluctuation induced transport may
fall in the same general class of problems as considered in section
IIIA. For example, earlier models of thermal ratchets driven by colored noise
in a small correlation time expansion (or in the unified colored noise
approximation for arbitrary time) become identical 
to a Brownian particle moving in an
inhomogeneous medium with space dependent diffusion coefficient [27].
The interesting idea of relative stability of states,
which affects current, in the presence of temperature 
nonuniformity, 
however, has not received the attention it deserves.

\par To conclude, we remark that we do not require the periodic
potential field of the system to be ratchetlike nor do we
require the fluctuating force to be a colored noise to obtain
macroscopic current. Instead, we put emphasis on the role of
inhomogeneity of the system to obtain unidirectional current. 
These inhomogeneities by themselves
(section III A) or with the help of external white noise
(section III B and III C) or with two temperature baths
at different temperatures, one of the bath being athermal (section IIID),
satisfy all the conditions to obtain
the macroscopic current. 
To summarise, in the present work, we have given a microscopic
basis to obtain unidirectional current in an inhomogeneous system. 
The present work may help to put the
problem of macroscopic unidirectional motion in
nonequilibrium systems on a more general footing.

\vfill
\eject

\vfill
\eject
\end{document}